\documentclass[iop]{emulateapj}
\usepackage{amssymb}
\usepackage{times}
\usepackage{epsfig}
\usepackage{graphics}

\usepackage{amsmath}

\newcommand{\beq}{\begin{eqnarray}}
\newcommand{\eeq}{\end{eqnarray}}
\newcommand{\bez}{\begin{eqnarray*}}
\newcommand{\eez}{\end{eqnarray*}}
\newcommand{\bc}{\begin{center}}
\newcommand{\ec}{\end{center}}

 \newcommand{\jsyn}{j_{\rm s}}
\newcommand{\npoel}{{n_{\rm e}}}
 
\newcommand{\Rph}{R_\star}
\newcommand{\mprot}{m_{\rm p}} 
\newcommand{\mneutr}{m_{\rm n}}
\newcommand{\sigmat}{\sigma_{\rm T}}
\newcommand{\sigman}{\sigma_{\rm n}}
   
\def\Beta{\beta}

\def\nneutr{n_{\rm n}}
\def\nprot{n}
\newcommand{\Te}{T_{\mathrm{e}}}
\newcommand{\taut}{\tau_{\mathrm{T}}} 
\newcommand{\taun}{\tau_{\mathrm{n}}} 
 
\def\cD{{\cal D}}

\def\Gamman{\Gamma_{\rm n}}

\def\zT{z_{\rm T}}

\def\Npm{n_{\pm}}
\def\jpm{j_{\pm}}
\def\kpm{\kappa_{\pm}}

\def\Es{E_{\rm s}}
\def\Ls{L_{\rm s}}
\def\nuB{\nu_{B}}

\def\eB{\varepsilon_{\rm B}}
\def\erad{\varepsilon_{\rm rad}}
\def\gammas{\gamma_{\rm s}}
\def\kappas{\kappa_{\rm s}}
\def\rphs{R_{\rm s}}
 
\def\Lrad{L_{\gamma}}

\def\TBB{T_{\rm BB}}
\def\LBB{L_{\rm BB}}
\def\Lp{L}
\def\Ln{L_{\rm n}}
\def\Gp{\Gamma}
\def\Gn{\Gamma_{\rm n}}
\def\Grel{\Gamma_{\rm rel}}
\def\texp{t_{\rm exp}}
\def\Epeak{E_{\rm peak}}
\def\Rn{R_{\rm n}}
\def\me{m_{\rm e}}

\def\M{{\cal M}}
\def\sss{\delta}
\def\alf{\alpha_f}

\begin{document}

\title{Gamma-ray bursts from magnetized collisionally-heated jets}

\shorttitle{GAMMA-RAY BURSTS FROM MAGNETIZED  COLLISIONALLY-HEATED JETS}
\shortauthors{VURM, BELOBORODOV, \& POUTANEN}

\author{Indrek Vurm,\altaffilmark{1,2,3} Andrei M. Beloborodov,\altaffilmark{4,5} and Juri Poutanen\altaffilmark{3}}

\affil{$^1$Racah Institute of Physics, Hebrew University of Jerusalem,  
Jerusalem 91904, Israel; indrek@phys.huji.ac.il \\ 
$^2$Tartu Observatory, T\~{o}ravere 61602, Tartumaa, Estonia\\ 
$^3$Astronomy Division, Department of Physics, P.O.Box 3000, 
90014 University of Oulu, Finland; juri.poutanen@oulu.fi \\
$^4$Physics Department and Columbia Astrophysics Laboratory, Columbia University, 538 West 120th Street New York, NY 10027; amb@phys.columbia.edu \\
$^5$Astro-Space Center of Lebedev Physical Institute, Profsojuznaja 84/32, Moscow 117810, Russia \\
}


\begin{abstract}
\noindent 
Jets producing gamma-ray bursts (GRBs) are likely to carry a neutron 
component that drifts with respect to the proton component. 
The neutron-proton collisions strongly heat the jet and generate 
electron-positron pairs. We investigate radiation produced by this heating 
using a new numerical code. Our results confirm the recent claim that 
collisional heating generates the observed Band-type spectrum of GRBs. 
We extend the model to study the effects of magnetic fields on the emitted 
spectrum. We find that the spectrum peak remains near 1~MeV for the entire 
range of the magnetization parameter $0<\eB<2$ that is explored in our 
simulations. The low-energy part of the spectrum softens with increasing 
$\eB$, and a visible soft excess appears in the keV band. The high-energy 
part of the spectrum extends well above the GeV range and can contribute to 
the prompt emission observed by {\it Fermi}/LAT. Overall, the radiation 
spectrum created by the collisional mechanism appears to agree with 
observations, with no fine-tuning of parameters.
\end{abstract}



\section{Introduction}	\label{chapter:GRB}

Gamma-ray bursts (GRBs) are produced by ultra-relativistic jets from 
short-lived and powerful energy sources, probably associated with black 
hole formation. Apart from the jet launching itself, a primary question 
of GRB theory concerns the emission mechanism of the burst: how does 
the jet emit the observed gamma-rays?

The jet must develop an ultra-relativistic speed if its initial thermal 
energy is much larger than its baryonic rest-mass energy.
The thermal energy is dominated by radiation and can be released 
at the photospheric radius $\Rph$ where the jet becomes transparent. 
For sufficiently clean (baryon-poor) thermal jets, transparency occurs
when radiation still carries most of the explosion energy.
Such a ``radiation-dominated'' GRB has a high radiative efficiency and 
a spectrum that peaks near 1~MeV \citep{Pacz86,Goodman86}. 
The shape of the spectrum in this case should be Planckian 
\citep[][hereafter B11]{B11}.

The observed emission indeed peaks near 1~MeV in most   bursts, however its 
spectrum is nonthermal, with an extended tail of high-energy emission.
This has two implications:
  (1) the GRB photosphere is rarely radiation-dominated, and 
(2) some form of dissipation operates in the GRB jets and 
greatly broadens their radiation spectra. The free energy available for 
dissipation may be kinetic or magnetic.
In particular, the relative motion of different parts of the jet can 
be dissipated \citep[e.g.][]{ReesMeszaros94,PaczXu1994}.
It leads to collisionless shocks as well as collisional dissipation.

Collisionless shocks are known as efficient accelerators of nonthermal 
particles in astrophysics. They offer a mechanism for generating 
nonthermal electrons in GRB jets, and their synchrotron emission
may be associated with observed $\gamma$-rays.
The idea became popular, however it faces difficulties. 
First, recent {\it ab initio} particle-in-cell simulations of collisionless
shocks \citep[e.g.][]{SironiSpitkovsky2009,SironiSpitkovsky2011}
do not support this scenario. 
Their results suggest that nonthermal electron acceleration by 
shocks is inefficient in GRB jets with expected transverse magnetic 
fields and magnetization parameter $\eB>10^{-3}$.
Secondly, the spectra of a large number of GRBs observed by {\it CGRO}/BATSE
and {\it Fermi}/GBM instruments are inconsistent with synchrotron emission, 
especially in the fast-cooling regime that is expected in the model 
\citep[e.g.][]{Preece00}. Some of the spectra 
show harder slopes than synchrotron emission can produce irrespective of 
the cooling regime. In addition, the synchrotron model does not predict the 
observed preferential
position of the spectral peak, which is always in the MeV range for
bright GRBs.
A recent discussion of 
the internal-shock synchrotron model can be found in \citet{Daigne11}.

One can consider other phenomenological models of emission from 
the optically thin region of the jet at $r>\Rph$. 
In particular, one can consider synchrotron self-Compton 
radiation from continually and uniformly heated plasma \citep{SP04,VP09}. 
In this scenario,
the optically thin plasma maintains a relativistic temperature, 
$kT_{\rm e}\gg m_{\rm e} c^2$, at which radiative cooling balances 
the heating, and keeps radiating as the jet expands.
While this model can produce hard spectra, it still does not explain
the preferential position of the observed spectrum peak. 
 It also conflicts with the recently observed high-energy spectra of GRBs
\citep[e.g.][]{Abdo_080916C,Zhang11}.

A likely resolution of these problems is that the bulk of GRB luminosity 
is generated below the photosphere at $r\lesssim\Rph$, rather than in the optically 
thin zone $r>\Rph$. 
Then multiple Compton scattering participates in the spectrum formation 
and can naturally produce the observed spectra. Generally,
any subphotospheric heating leads to Comptonization of the (initially 
thermal) photons, and then radiation released at the photosphere $\Rph$ 
looks as nonthermal while the peak of its spectrum remains near 1~MeV
(\citealt{Thompson94,MR00b,Giannios06,Peer06,B10}, hereafter B10).
A variety of Comptonized spectra may be generated, depending on the 
assumptions of the model.
Further progress can only be made when the physical mechanism of energy dissipation  
is understood. It could be thermal or nonthermal, and its radial dependence 
is important for the emerging spectrum.

Three heating mechanisms can operate in the subphotospheric region. 
Two of them (internal shocks and magnetic dissipation) rely of collisionless 
plasma processes, whose details remain uncertain because of their complexity.
The third mechanism (collisional heating) is straightforward and can be 
modeled from first principles. It was recently proposed that collisional 
heating shapes the GRB spectra (B10). It will be the focus of the present 
paper.

Collisional heating is particularly strong (and inevitable) in jets that 
carry free neutrons 
\citep{Derishev99,BahcallMeszaros2000,Fuller2000,MR00a,Rossi06,KG07}.
Before the complete decoupling of neutrons and protons into two 
non-interacting components, the fading $n$-$p$ collisions heat the 
protons to a mildly relativistic temperature. Importantly, the $n$-$p$ 
collisions also create $e^\pm$ plasma that efficiently radiates its energy.
The $e^\pm$ plasma is cooled by radiative losses and 
heated by Coulomb collisions with hot protons. The rates of all these 
processes are well known and their effect on the jet and radiation can be 
calculated without additional assumptions (B10). In particular, one can 
show that the electron distribution function has two parts, thermal and 
nonthermal, and both Comptonize the radiation carried by the jet. 

Three features of collisional heating make it a promising emission 
mechanism:
(1) It peaks at radii $r\sim 0.1\Rph$, not too early and not too late 
to produce a bright photospheric emission.
(2) It has a high radiative efficiency, comparable to 50\%.
(3) The radiation spectrum produced by collisional heating can be 
accurately calculated from first principles. 
The striking result reported in B10 is that the spectrum 
has the Band-type shape with the high-energy photon index $\beta\sim -2.5$, 
which is consistent with observations. 

The formation of the radiation spectrum in a heated jet is a nonlinear problem 
that requires self-consistent calculation of the plasma and radiation 
behavior. This problem was solved in B10 using a Monte-Carlo radiative 
transfer code in combination with an iterative technique.
Our present work has two goals:

First,  to provide an independent check of the result of B10, we calculate 
the emitted spectrum using a
different method: we solve the kinetic equations for the plasma and 
radiation in the jet. We use the numerical code described in 
\citet[][hereafter VP09]{VP09}. 
We develop a new version of the code to adapt it to GRB jets, calculate 
the emerging radiation spectrum, and compare the results with those of B10. 

Secondly, the numerical models in B10 were limited to weakly magnetized flows,
$\eB \ll 1$, and neglected synchrotron emission. In the present paper,
our models include synchrotron emission and self-absorption, and we  
explore the spectra produced by magnetized neutron-loaded jets for a range 
of $0<\eB <2$.

The paper is organized as follows. Section~\ref{GRB:Simulsetup} describes
the physical model and the setup of the simulations. Section~\ref{sec:method}  
describes our method of calculations. The results are presented in 
Sections~\ref{GRB:results1} and \ref{GRB:results2} and discussed in 
Section~\ref{GRB:conclusions}.

\vspace{5mm}

\section{Physical model and simulation setup}
\label{GRB:Simulsetup}

The central engines of GRBs must be compact and hot objects, which
are almost certainly neutron rich. The high density and temperature
inevitably leads to $\beta$-equilibrium 
\citep{Imshennik67,Derishev99,Beloborodov03}.
For all plausible parameters of the central engine, the equilibrium 
establishes a high fraction of free neutrons in the baryonic matter 
(this feature is seen, for example, in the accretion-disk model for 
GRBs, see \citealt{Beloborodov08} for a review).
Therefore, neutrons are generally expected in GRB jets.

The presence of neutrons creates perfect conditions for strong 
collisional heating. In any variable baryonic jet, the 
inter-penetrating neutron and proton components inevitably develop 
relative motions (see B10 and references therein).
The two components acquire different Lorentz factors $\Gn$ and $\Gp$, and 
(rare) collisions between neutrons and protons dissipate enormous energy. 
The dissipation is strongest in those parts of the jet where $\Gp/\Gn$ is 
highest. 

The variable jet consists of many causally disconnected shells.
Different shells are accelerated to different Lorentz factors and emit 
different radiation, which is consistent with observed variability.
In our simulations, we focus on the heating history and radiation of 
one shell. The shell is launched from the central source and accelerated 
at the expense of its internal energy until
it enters the coasting matter-dominated phase of expansion.
At radius $R_{\rm n}$, the timescale for collisions between the neutron 
and proton components exceeds the expansion timescale of the jet, 
and protons start to migrate relative to neutrons, forming a compound flow
with $\Gp\neq\Gn$. At this point, strong collisional heating begins and the 
radiation spectrum starts deviating from a blackbody.
Our simulation starts at $R_{\rm n}$ and follows this evolution.
We assume that $\Gp$ and $\Gn$ remain approximately constant till the end 
of the simulation (see B10 for discussion of this approximation).
Note that migration of neutrons between different shells of a
strongly variable jet can create compound flows with $\Gamma\gg\Gn$.

$R_{\rm n}$ may be defined as the radius where the neutron flow becomes 
``optically thin'' to collisions with protons. The corresponding  
``optical depth'' for this process at radius $r$ is given by
\beq
 \taun = \frac{\nneutr\sigman r}{\Gamman}
       = \frac{\Ln\sigman}{4\pi \mneutr c^3 r \Gn^3}=\frac{R_{\rm n}}{r},
\eeq
where $\nneutr$ is the comoving number density of the neutron flow,
$\Ln = 4\pi \mneutr c^3 r^2  \Gn^2 \nneutr$ is its kinetic luminosity
(isotropic equivalent) and
$\sigman \sim 3 \times 10^{-26} \, {\rm cm}^2$ is the effective 
cross-section for nuclear collisions. 
Radius $R_{\rm n}$ is defined so that $\taun(R_{\rm n})=1$,
\beq
\label{dissip:rn}
  R_{\rm n} =  \frac{\Ln\sigman}{4\pi \mneutr c^3 \Gn^3}.
\eeq

We aim to calculate the evolution of electron/positron and photon 
distribution functions in the heated jet.  The photon spectrum is 
controlled by electrons (and positrons) via Compton scattering and 
synchrotron emission. Therefore, the key ingredient of any GRB model
is how energy is injected into electrons. For the model studied in this 
paper, the electron heating is unambiguously determined by the collisional 
processes as described below (see B10 for detailed discussion).

The electron heating comes from baryons, in two forms:
\begin{enumerate}
\item Nuclear $n$-$p$ collisions generate $e^\pm$ pairs whose energy in the comoving frame of the jet is comparable
  to the pion rest-mass, $\gamma\me c^2\sim m_\pi c^2\approx 140$~MeV. 
The injection rate of these energetic particles is proportional to the 
rate of nuclear collisions, $\dot{n}_{\rm coll}=c\sigman\nprot\nneutr\Grel$,
where $\Grel\approx\frac{1}{2}(\Gp/\Gn+\Gn/\Gp)$ is the relative Lorentz factor 
of collisions. The $e^\pm$ injection rate is given by (B10),
\begin{equation} \label{eq:nonth}
 \dot{n}_\pm^{\rm inj}\approx\frac{1}{4}\,
       \frac{\Grel \mprot}{\gamma_0 m_{\rm e}}\,\dot{n}_{\rm coll}
  \approx \frac{3}{8}\,\frac{\Gp}{\Gn}\,\taun\,\frac{\nprot}{\texp}. 
\end{equation}
Here $\nprot$ is the proton number density in the plasma rest frame and
\begin{equation}
   \texp\equiv\frac{r}{c\Gp}
      \label{dissip:texp}
\end{equation}
is the expansion timescale of the plasma, measured in its rest frame.
The $e^\pm$ injection peaks near the Lorentz factor 
$\gamma_0\approx m_\pi/m_{\rm e}\approx 300$.
We will approximate it by a Gaussian distribution centered at $\gamma_0=300$.
These energetic particles experience rapid radiative cooling and 
join the thermal $e^\pm$ population.

\item  The $e^\pm$ population receives energy from protons via Coulomb 
collisions.
Note that protons have at least mildly relativistic random velocities
--- they are stirred by $n$-$p$ collisions. 
The dominant majority of $e^\pm$ pairs are kept at a much lower 
temperature, because of Compton cooling.
Coulomb collisions  drain the energy from protons to  $e^\pm$ with
the following rate \citep[e.g.][]{GS64}
\begin{equation}   \label{eq:th}
  \dot{Q}_{\rm th} = \frac{3}{2}\ln\Lambda\,\frac{\sigmat\,m_{\rm e} c^3\,\npoel\nprot}
            {\beta_{\rm p} }
          \approx 0.02 \taut \,\frac{\nprot \mprot c^2}{\texp},
\end{equation}
where $\ln\Lambda$ is the Coulomb logarithm, $\beta_{\rm p} \sim 1$ 
is the random velocity of the protons stirred by nuclear collisions, 
$\npoel=n_-+n_+$ is the number density of electrons and positrons, and
\begin{equation} \label{eq:taut}
  \taut=\frac{\npoel\sigmat r}{\Gp}
\end{equation}
is a characteristic Thomson optical depth of the plasma, as seen by 
photons at radius $r$.
The pair density $\npoel$ is controlled by the process of $e^\pm$ injection (Equation~\ref{eq:nonth}),
the subsequent pair-photon cascade in the radiation field, and $e^\pm$ annihilation, all of which
must be self-consistently calculated. 
\end{enumerate}

Equations~(\ref{eq:nonth}) and (\ref{eq:th}) determine the heating 
history of the expanding plasma shell and eventually the 
radiation spectrum emitted by this shell. The shell itself is a part of 
the (variable) jet. It has the following parameters:
\begin{enumerate}
\item 
Lorentz factor and kinetic luminosity ($\Gp$ and $\Lp$ for the 
plasma component and $\Gn$ and $\Ln$ for the neutron component).
Note that the characteristic radius $R_{\rm n}$ for a given shell is 
determined by its $\Gn$ and $\Ln$ according to Equation~(\ref{dissip:rn}).

\item 
Temperature of the thermal radiation at $R_{\rm n}$, prior to the onset of 
collisional heating, $\TBB(R_{\rm n})$. The corresponding luminosity is
$\LBB=(4/3)\,c\,a\TBB^4\Gp^2 4\pi R_{\rm n}^2$ (here $a$ is the radiation 
density constant). This parameter depends on $\Lp$, $\Gp$, and the radius at 
the base of the jet $r_0$ where our shell starts to accelerate. We fix 
$r_0=10^7$~cm in our numerical models and calculate the corresponding 
$\TBB(R_{\rm n})$ using the standard model of adiabatically cooled outflow 
between $r_0$ and $R_{\rm n}$ \citep[see e.g.][]{Pacz90,Piran93}. 

\item 
Magnetization $\eB$. It is defined as the ratio of the magnetic energy 
density $U_{\rm B}=B^2/8\pi$ to the proper energy density of the plasma flow
$U=\Lp/4\pi cr^2\Gp^2$ 
  (which includes rest-mass energy).
Both $U_{\rm B}$ and $U$ are measured in the comoving 
frame of the plasma. We assume that the 
magnetic field is advected from the central engine. Then the field must 
be transverse to the (radial) velocity\footnote{
    The radial component $B_r$ is suppressed as $r^{-2}$ in the 
    expanding plasma and can be neglected.}
and scale with radius as $B\propto r^{-1}$
(assuming $\Gamma\approx const$).
In this case $\eB=U_{\rm B}/U\approx const$, i.e. does not change with radius.
\end{enumerate}


\section{Method of calculations}
\label{sec:method}

For our calculations we use  a new version of
the numerical code developed by VP09.
The code is designed to model the coupled evolution of 
 the heated outflow and the radiation it carries.
The evolution is tracked by solving the time-dependent kinetic equations 
for the particles and photons. The included interactions are Compton 
scattering, cyclo-synchrotron emission and absorption, photon-photon pair 
production and annihilation, and Coulomb collisions. 
We use the exact cross-sections or rates for all these processes.

The original version of the code had one significant limitation:
it used the ``one-zone'' or ``leaking-box'' approximation. 
This approximation pictures a uniform plasma cloud of an optical 
depth $\taut$ with isotropic populations of particles and photons.
It treats the loss of photons from the cloud using an escape probability 
instead of accurate calculations of radiation diffusion through the cloud 
(VP09). 

The leaking-box picture is not good for GRB jets. 
In contrast to static sources, where radiation escapes the source 
on timescale $\sim\taut R/c$, GRB radiation remains embedded in the 
relativistic jet 
  at all radii of interest. It evolves with radius according to the 
  radiative transfer equation, which is in serious conflict with the 
  one-zone approximation. For example, the true angular distribution of 
  radiation in the jet comoving frame is far from being isotropic, even 
  in the subphotospheric region (B11).
The isotropic approximation of the one-zone model becomes invalid
when the optical depth $\taut$ decreases below $\sim 10$. 
 Furthermore,
the large free path of photons near the photosphere leads to 
mixing of radiation emitted by different parts of the photospheric 
region with different Doppler shifts. 
  This mixing has a significant effect on 
the local photon spectrum and the spectrum received 
by a distant observer. 

For these reasons the simplified one-zone treatment has to be abandoned in 
favor of proper radiative transfer calculations. The kinetic equation for 
photons in VP09 is replaced by the transfer equation
as described below.

\subsection{Radiative transfer}
\label{sec:rad_transf}

The equation of radiative transfer in an ultra-relativistic,
matter-dominated outflow reads (B11)
\begin{align}
 \frac{\partial I_\nu}{\partial \ln r } & =
      (1-\mu)\,\left(\frac{\partial I_\nu}{\partial\ln\nu}-3I_\nu\right) 
    - (1-\mu^2)\,\frac{\partial I_\nu}{\partial\mu} \nonumber \\
 &  + \frac{r\,(j_{\nu} - \kappa_\nu I_\nu)}{\Gamma\,(1+\mu)}. 
\label{method:ph_eq}
\end{align}
Here $I_{\nu}$ is the specific intensity, $\nu$ is the photon frequency, 
$\mu= \cos{\theta}$, $\theta$ is the angle relative to the radial 
direction, and $r$ is the distance from the central source. 
The emission and absorption coefficients $j_{\nu}$ and $\kappa_{\nu}$ 
represent all the interactions of radiation with plasma
as well as with radiation itself.
All quantities (except $r$ and $\Gamma$) are measured in the
rest frame of the outflow.

Equation~(\ref{method:ph_eq}) and the formulation of the 
ultra-relativistic transfer problem are discussed in detail in B11. 
The problem simplifies because essentially all photons flow toward larger 
$r$, and hence only inner boundary conditions need to be specified. 
The radiative transfer then takes the form of an initial-value problem. 
One can think of it as the evolution of $I_\nu$ with $r$ or, 
equivalently, with the jet comoving time $t$.
Equation~(\ref{method:ph_eq}) is similar to the kinetic equation 
for the evolution of the photon distribution function (VP09), 
with additional terms due to the angular dependence. 
 
The inner boundary condition for $I_\nu$ is the blackbody radiation of 
a given temperature. We track the evolution of radiation with $r$
consistently with the plasma evolution, which is described in Section~\ref{sec:kin_plasma}.
In the case of a passively cooling jet (no heating), the transfer equation 
reproduces the adiabatic cooling of photons in the subphotospheric 
region. Heating changes the state of the plasma and greatly affects the 
source function $S_\nu=j_\nu/\kappa_\nu$ that governs the radiative transfer. 

Our numerical solution of Equation~(\ref{method:ph_eq}) will use one 
approximation to reduce the computational time: when calculating $j_\nu$ 
and $\kappa_\nu$ we pretend that radiation is isotropic, using 
only the zeroth moment of the actual angular distribution
of photons.
In particular, the scattering rates are averaged over angles.
The approximation is known to be quite successful in classical transfer 
problems \citep[e.g.][]{Ch60}. 
It is also reasonable for GRB 
outflows if the bulk of scattering events are non-relativistic in 
the plasma rest frame (B11).
Its error becomes more significant at high photon energies, comparable 
to $m_{\rm e}c^2$ in the plasma frame. 
This approximation allows us to use the numerical tools for $j_\nu$
and $\kappa_\nu$ developed in VP09. A detailed discussion of 
the angle-averaged $j_\nu$ and $\kappa_\nu$ and their numerical treatment 
is found in that work.

Radiation that will be received by a distant observer is determined by 
$I_\nu$ at a large enough radius, where photons stream almost freely at 
{\it all} relevant $\nu$ (note that this radius is much larger than $\Rph$).
Using $1-\Gamma^{-2}\approx 1$, we define Doppler factor as 
$\cD = \Gamma(1+\mu)$. The spectral distribution of the observed (isotropic 
equivalent) luminosity $\Lrad$ is given by the Lorentz transformation 
of $I_\nu\,d\mu$
to the observer frame and integration over angles,
\begin{align}
  \Lrad(E) \equiv \frac{d\Lrad}{dE} 
          = \frac{8\pi^2}{h} R^2 \int_{-1}^1 \cD(\mu)\,I_\nu(\mu)\,d\mu,
\label{eq:Lg}
\end{align}
where $E=\cD h\nu$ is the observed photon energy and $h$ is the Planck
constant. We emphasize that the simple relation~(\ref{eq:Lg}) is valid 
only at large radii, after all transfer effects on $I_\nu$ have 
already
been calculated according to Equation~(\ref{method:ph_eq}). 

Finally, we note that Equation (\ref{method:ph_eq}) was formally derived
for steady outflows, however it is also applicable to strongly variable, 
non-uniform jets (see B11 and Appendix \ref{append:RTE}).
The basic reason for this is that radiation moves together with 
the ultra-relativistic plasma flow.
The photon diffusion in the plasma rest frame is limited to scales 
$\delta\sim ct\sim r/\Gamma$, which are much smaller than $r$. 
  When viewed in the fixed observer frame, the scale of diffusion in the 
  radial direction is additionally compressed by the factor of $\Gamma^{-1}$.
The transfer occurs in a thin shell (or ``pancake'') of the jet 
as if it were part of a steady outflow.

\subsection{Kinetic equation for plasma}
\label{sec:kin_plasma}

The comoving-frame kinetic equation for electrons and positrons can be 
written as
\begin{align}
  \frac{\partial \Npm(p)}{\partial t} &
  = -\frac{\partial}{\partial p} \left[ \dot{p}\, \Npm(p)
    - \frac{1}{2} \frac{\partial}{\partial \gamma} \left( D \Npm(p) \right)
  \right]							\nonumber \\
         & + \jpm - \kpm \Npm(p) - \frac{2}{t}\,\Npm(p).
\label{method:pair_eq}
\end{align}
Here $p=\sqrt{\gamma^2-1}$ is the electron/positron momentum in units of $m_ec$,
$\Npm(p)$ are the distribution functions of positrons and electrons,
and $t=r/c\Gamma$ is the proper time of the flow. 
When calculating $\dot{p}$ we take into account
Coulomb heating by protons (Equation~\ref{eq:th}) and adiabatic 
cooling (Equation~\ref{Implem:dotg} below).
The term $\dot{p}$
also includes the usual contributions from 
Compton scattering (in the Thomson regime), 
synchrotron emission and self-absorption, and Coulomb collisions between 
$e^\pm$, as described in VP09. The diffusion coefficient $D$ governs 
particle thermalization at low energies due to synchrotron processes  and 
Coulomb collisions between $e^\pm$ (VP09).
The source $\jpm$ includes the injection of high-energy pairs by
inelastic nuclear collisions as described in Section~\ref{GRB:Simulsetup}.
The source and sink terms $\jpm$ and $\kpm$ also contain contributions from
photon-photon pair production, pair annihilation, and Compton scattering 
(in the Klein-Nishina regime). Our code calculates all these processes 
using the exact cross-sections, as described in detail in VP09. 
Finally, the last term in Equation~(\ref{method:pair_eq}) takes into account 
the dilution of particle densities due to the two-dimensional (sideways) 
expansion of the coasting flow.

The contribution of adiabatic cooling to $\dot{p}$ is given by 
\begin{align}\label{Implem:dotg}
  \dot{p}_{\rm ad} = - \frac{2}{3} \, \frac{p}{t}.
\end{align}
This is easy to verify by considering adiabatic expansion of
${\cal N}_e$ monoenergetic electrons occupying a volume $V$.
Their pressure $P_{\rm e} = ({\cal N}_{\rm e}/V)(p^2/3\gamma)m_ec^2$
determines the adiabatic cooling rate 
${\cal N}_{\rm e} m_ec^2 d\gamma = -P_{\rm e} \,dV$. 
The coasting outflow expands in two dimensions, $V\propto r^2\propto t^2$.
Using $dp=(\gamma/p)d\gamma$, one gets Equation~(\ref{Implem:dotg}).

In our numerical simulations, the electron/positron kinetic equation
(\ref{method:pair_eq}) is discretized on a grid of particle momenta 
and solved simultaneously with the radiative transfer equation
(\ref{method:ph_eq}).
The grid of dimensionless momentum $p$
extends from $10^{-4}$ to $10^4$, with 25 points per decade.
The photon-energy grid spans 15 decades from $h\nu/m_{\rm e} c^2 = 10^{-11}$ 
to $10^4$ (comoving frame), with approximately 13 gridpoints per decade.
The lower boundary is set so low to properly simulate the effects of
synchrotron self-absorption in mildly relativistic pairs in a weak magnetic
field. The photon angular grid is uniform in $\theta$ and has 40 grid points.


\section{Non-magnetized outflows}
\label{GRB:results1}

We begin with the non-magnetized model that can be directly compared 
with the results in B10. The model has the following parameters:
proton flow luminosity $L = 10^{52}$~erg~s$^{-1}$, neutron flow 
luminosity $\Ln = 2 \times 10^{51}$~erg~s$^{-1}$,
Lorentz factor of the proton flow (baryon loading) $\Gamma = 600$,
Lorentz factor of the neutron flow $\Gamman = 100$,
initial radius of the flow $r_0 = 10^7$ cm.
The starting radius of the simulations is determined by 
Equation~(\ref{dissip:rn}), which gives $R_{\rm n} = 10^{11}$ cm.
The comoving temperature of the blackbody radiation field at $R_{\rm n}$
is found from the passively cooling outflow model 
at $r<R_{\rm n}$, which gives  $k\TBB(R_{\rm n}) \approx 0.5$ keV.

\subsection{Electron/positron distribution}
\label{sec:ele_dist_nonmag}

At the radius $R_{\rm n}$, where the dissipation begins,
the injected high-energy pairs start to upscatter the thermal photons
to energies up to $\sim 100 \me c^2$ in the rest-frame of the jet.
The optical depth for photon-photon pair production seen by the high-energy 
photons is initially small, as the
target photons are confined to
the blackbody component that is below the threshold for 
the reaction. However, this soon changes as more and more photons 
are upscattered from the thermal distribution. Then the jet becomes 
optically thick to pair production and a cascade develops. Soon the 
annihilation balance is established, in which the cooled $e^\pm$ pairs 
annihilate at the same rate as the new pairs are created by the cascade.

\begin{figure}
\plotone{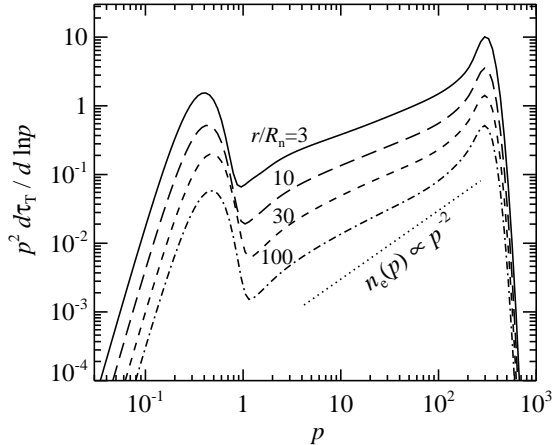}
\caption{
  Momentum distribution function for electrons and positrons in the 
  non-magnetized jet at different radii:
  $r/R_{\rm n}=3$, 10, 30, and 100 (solid, long-dashed, dashed and 
  dot-dashed curves, respectively).
  The distribution is normalized so that it shows the optical depth $\taut$, 
  which is related to density $\npoel =n_++n_-$ by Equation~(\ref{eq:taut}). 
  The dotted line indicates the slope of a nonthermal distribution that 
  would be obtained if the injected $e^\pm$ did not produce a cascade
  and were Compton-cooled in the Thomson regime.
}
\label{fig:ele_distr_nonmag}
\end{figure}

The continual cascade quickly reaches a steady state, on a timescale
shorter than the jet expansion timescale. Then a quasi-steady 
$e^\pm$ distribution is maintained, which gradually changes with radius 
(Figure~\ref{fig:ele_distr_nonmag}).
The distribution has two parts, thermal and nonthermal (cf. B10).
The nonthermal tail consists of continually injected, fast-cooling
$e^\pm$ pairs. If the injected $e^\pm$ generated no cascade and cooled in 
Thomson regime at all $p$, the nonthermal tail would have a power-law 
form $\npoel(p)\propto p^{-2}$.
The actual distribution is affected by the Klein-Nishina effects in
Compton cooling (which are significant at $p>50$) and the large 
number of secondary pairs produced in the cascade.

The low-energy peak of the $e^\pm$ distribution maintains a Maxwellian shape
due to Coulomb collisions between $e^\pm$ particles. This thermalized 
population is continually heated by Coulomb collisions with protons and 
maintains the equilibrium temperature that is determined by the balance 
between Coulomb heating and Compton cooling. The resulting electron 
temperature is close to 15~keV and remarkably stable throughout the 
dissipation region (see Figure~\ref{fig:ele_temp} below). Its weak radial 
dependence is approximately described by equation~(32) in B10. 
 
Shortly after the dissipation starts, the scattering opacity of the flow 
becomes dominated by the created $e^\pm$ pairs.
The Thomson optical depth is regulated to $\taut\approx 20$ near $R_{\rm n}$ 
and then decreases as $R_{\rm n}/r$ (see Figure~\ref{fig:ele_distr_nonmag}),  
in excellent agreement with B10. 
This places the photospheric radius at $\Rph\sim 20R_{\rm n}$. 

Since $\taut \propto \npoel r \propto r^{-1}$, Equation~(\ref{eq:th}) 
shows that the   Coulomb heating peaks near $R_{\rm n}$,
and the fraction of the flow energy dissipated in one dynamical time 
at larger radii scales as $r^{-1}$. The nonthermal energy injection 
has the same $r^{-1}$ dependence. 
Thus, the ratio of thermal and nonthermal heating rates remains nearly 
constant along the flow. This ratio is close to unity.

Most of the kinetic energy of the flow is dissipated (and re-dissipated) 
by collisional heating. One should keep in mind though that the heat 
given to plasma and radiation in the subphotospheric region is 
continually degraded by adiabatic cooling. As a result, the actual 
luminosity released at the photosphere is smaller by a factor of 1/2 
compared with the total heat deposited in the flow (see section~4.6 in B10).

\begin{figure}
\plotone{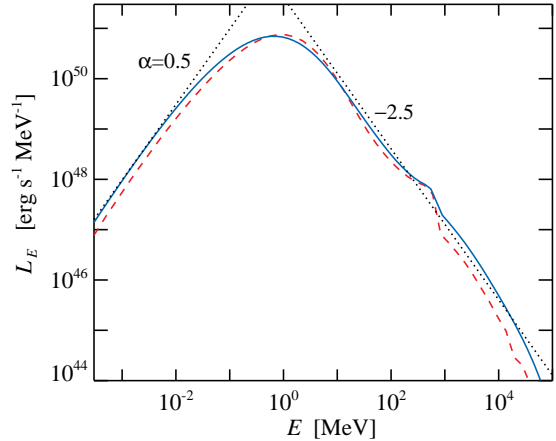}
\caption{Comparison of spectra obtained by kinetic (solid line) and 
Monte Carlo (dashed line) simulations for the non-magnetized jet.
Model parameters: proton flow luminosity $L = 10^{52}$ erg s$^{-1}$, 
neutron flow luminosity $\Ln = 2 \times10^{51}$ erg s$^{-1}$,
Lorentz factor of the proton flow $\Gamma = 600$,
Lorentz factor of the neutron flow $\Gamman = 100$.
The dotted straight lines correspond to
photon indices $0.5$ and $-2.5$. 
Note that the calculated spectra are not corrected for the 
cosmological redshift $z$. All photons should be redshifted by the 
factor $(1+z)^{-1}$, which is $\sim 1/3$ for a typical GRB.
}
\label{fig:GRB_nm1}
\end{figure}

\subsection{Radiation spectrum}
\label{sec:rad_spe_nonmag}

The spectrum that should be received by a distant observer from the 
collisionally heated jet is shown in Figure~\ref{fig:GRB_nm1}.
For comparison, we also show the result of the Monte-Carlo simulation 
in B10. The two spectra are very similar.  

The emitted spectrum peaks near 1~MeV. Radiation below the peak is made 
of blackbody photons advected from the central source and released at the 
photosphere. Note that the low-energy photon index 
$\alpha\approx 0.4-0.5$ is significantly different from the 
Rayleigh-Jeans $\alpha=1$. This softening is a result of the
superposition of emissions from different parts of the photosphere having 
different Doppler factors, i.e. effectively a multi-temperature spectrum
is observed below the peak (B10). The spectrum between 1 and 20 MeV
is mainly shaped by thermal Comptonization by Coulomb-heated pairs.
The thermal pairs upscatter photons to energies up to
$E\sim 2\Gamma k \Te\sim 20$~MeV in the observer frame, 
where $k\Te\sim 15$~keV is the electron temperature in the comoving frame.

Above 20~MeV, the observed spectrum is generated by inverse Compton scattering by 
the nonthermal pairs.
Note that there is no dip between the thermal and nonthermal
Comptonization
components. 
Two reasons contribute to this remarkable smoothness of the spectral shape.
First, the energy budgets of thermal and nonthermal Comptonization are 
comparable, and the two components have comparable luminosities at 10--20~MeV. 
Secondly, most of the photons above 20~MeV undergo several scattering events 
on thermal pairs before escaping. They experience significant Compton 
downscattering, which has an overall smoothing effect on the spectrum.
The only distinct feature in
the predicted high-energy spectrum is a broad 
annihilation line on top of the smooth continuum just below 1~GeV.
The line is produced by the
cooled $e^\pm$ pairs (photon energy $E\sim 1$~GeV in the observer frame 
approximately corresponds to $h\nu\sim \me c^2$ in the jet frame). 

The observed spectrum at energies $E > 1$~GeV is affected by absorption 
due to  photon-photon ($\gamma$-$\gamma$) pair production. The jet is 
optically thick to GeV photons close to $\Rn$ where most of the heating 
occurs, and all high-energy upscattered photons are quickly absorbed, 
creating secondary $e^\pm$ pairs. Thus, the developing pair-photon cascade 
is initially in the saturated regime.

The photon emissivity $j_\nu$ produced by the saturated cascade may be 
estimated analytically \citep{Sve87}. An important parameter of the cascade 
is $\zT = (2/3) \gamma_0 x_0$, where $\gamma_0$ is the injection Lorentz 
factor of the primary nonthermal pairs and $x_0=h\nu_0/m_{\rm e} c^2$ is 
the typical energy of soft photons. 
The quantity $\zT$ determines whether the cascade takes place in the Thomson  
or Klein-Nishina regime ($\zT \le 0.5$ or $ > 0.5$, respectively), 
the number of generations of secondary pairs etc.
In our case $x_0 \sim 1/200$ and $\gamma_0 = 300$, so $\zT \sim 1$, 
i.e. the cascade starts in the Klein-Nishina regime at high energies
and proceeds in the Thomson regime at lower energies. There are several 
generations of secondary pairs and photons, leading to a smooth overall 
spectrum. For the cascade with $\zT\sim 1$ the analytic solution predicts a 
flat emissivity of high-energy photons, $\nu j_\nu\approx const$ \citep{Sve87}.
The radiation intensity inside an opaque source is given by 
$I_\nu=j_\nu/\kappa_{\nu,\gamma\gamma}$. Here the absorption 
coefficient $\kappa_{\nu,\gamma\gamma} \propto \nu^{-\beta-1}$  
and $\beta$ is the photon index of the target radiation (typically photons 
of multi-MeV energy in the observer frame). Thus, the high-energy spectrum 
$I_\nu \propto \nu^\beta$ is maintained inside the opaque jet.

The escaping multi-GeV emission may be estimated by considering the 
evolution of $\gamma$-$\gamma$ opacity with radius (B10).
The rate of nuclear collisions fades at large radii, but it still generates 
a significant $e^\pm$ cascade. Since the $\gamma$-$\gamma$ optical depth  
decreases with $r$, the multi-GeV photons produced in the outer region 
have a chance to escape. The radius of $\gamma$-$\gamma$ transparency at 
energy $E$ scales approximately as 
$R_{\gamma\gamma}\propto E^{-\beta-1}$, and the power of the cascade
scales as $r^{-1}$. This leads to $\Lrad(E) \propto E^\beta$
(photon index $\beta-1$).\footnote{
    Incidentally, the high-energy radiation inside the opaque source 
    $I_\nu=j_\nu/\kappa_{\nu,\gamma\gamma}$ also has the photon index 
    $\beta-1$. This coincidence is a result of the particular scaling 
    $j_\nu\propto r^{-4}$ and $\kappa_{\nu,\gamma\gamma}\propto r^{-2}$ 
    in the collisionally heated jet.}   
This rough estimate suggests the change in photon index above a few GeV 
from $\beta$ to $\beta-1$. A similar steepening of the spectrum is observed 
in Figure~\ref{fig:GRB_nm1}.
Note that the high-energy radiation escaping from the collisionally heated 
jet extends far beyond 1~GeV, up to 100~GeV.
 
Finally, note that our spectrum
deviates from that of B10 
more significantly
in the GeV range. This is likely caused by our angle-averaged approximation for 
$j_\nu$ and $\kappa_\nu$ (Section~\ref{sec:rad_transf}). Anisotropy effects 
are particularly significant at the high-energy end of the spectrum that 
forms at large radii. At $r\gg\Rph$ the soft photon field is strongly 
collimated, and the scattering of nearly radial photons by relativistic 
pairs is preferentially backward. 
The backward-scattered photons see a higher $\gamma$-$\gamma$ optical depth 
and less of them can get out. The anisotropy of $\gamma$-$\gamma$ opacity is 
further enhanced by the very same collimation of the soft radiation field.
These effects create particularly strong angular dependence of $j_\nu$ 
and $\kappa_\nu$ at high energies, which is missed by our angle-averaged
approximation.


\section{Magnetized outflows}
\label{GRB:results2}

Magnetization  of the jet is described by the parameter
\begin{equation}
\label{eq:eB}
  \eB =\frac{L_{\rm B}}{L}=\frac{4\pi r^2\, c\, \Gamma^2\,U_{\rm B}}{\Lp},
\end{equation}
where $U_{\rm B}=B^2/8\pi$ is the magnetic energy measured in the rest frame of 
the outflowing plasma. The quantity $\eB$ is the ratio of the Poynting 
flux $L_B$ (measured in the static frame) to the kinetic luminosity of the 
outflow $\Lp$. 
We discuss below
the GRB model with the same parameters as in 
Section~\ref{GRB:results1} and investigate how the results change
with increasing $\eB$.
We have calculated a set 
of models with $\eB = 10^{-3}$, $10^{-2}$, $0.1$, $0.5$ and $2$.
The corresponding ratios of (comoving) magnetic and radiation energy 
densities at the start of simulations are 
$U_{\rm B}(R_{\rm n})/U_\gamma(R_{\rm n}) = 0.006$, $0.06$, $0.6$, $3$ and 
$12$. 
The results of our calculations are summarized in Table~\ref{tbl-1} and described in detail below.

\begin{table}
\begin{center}
\caption{Results of simulations. \label{tbl-1}}
\begin{tabular}{lcccrlll}
\tableline\tableline
$\eB$\tablenotemark{a}  &  $E_{\rm peak}$\tablenotemark{b}  &  $\alpha$\tablenotemark{c}  &  $\erad$\tablenotemark{d}  &
$\Rph/R_{\rm n}$\tablenotemark{e}  &  $Y$\tablenotemark{f}  &  $\dot{Q}_{\rm th}/\dot{Q}_{\rm nth}$\tablenotemark{g}	&	$kT_{\rm e}$\tablenotemark{h} \\
	&	(MeV)	&		&		&		&		&		&	(keV)	\\
\tableline
0			&	2.9	&	$-$0.6		&	0.46	&	16.5	&	0.19	&	0.94			&	15.0	\\
$10^{-3}$	&	2.5	&	$-$0.8		&	0.46	&	16.1	&	0.18	&	0.92			&	14.7	\\
$10^{-2}$	&	1.7	&	$-$1.2		&	0.45	&	14.2	&	0.14	&	0.79			&	13.8	\\
0.1			&	1.2	&	$-$1.4		&	0.46	&	8.8		&	0.058	&	0.32			&	13.1	\\
0.5			&	1.3	&	$-$1.4		&	0.52	&	2.3		&	0.014	&	0.10			&	13.1	\\
2.0			&	1.4	&	$-$1.3		&	0.55	&     0.5		&	0.005	&	0.05			&	14.4	\\
\tableline
\end{tabular}
\tablenotetext{1}{Magnetization defined in Equation~(\ref{eq:eB}).}
\tablenotetext{2}{Peak energy of the $EL_E$ spectrum.}
\tablenotetext{3}{Photon index in the 100--500 keV range.}
\tablenotetext{4}{Radiative efficiency $\erad = \Lrad/L$.}
\tablenotetext{5}{
Radius of the Thomson photosphere $\Rph$ relative 
to radius $R_{\rm n} = 10^{11}$ cm where the dissipation starts. 
In the case of $\eB=2$, the jet remains optically thin 
to scattering
throughout the collisionally heated region ($\Rph<\Rn$).}
\tablenotetext{6}{Pair yield $Y = {\cal M} /\gamma_0$,
where ${\cal M}$ is the secondary pair multiplicity and $\gamma_0 = 300$ is 
the Lorenz factor of injected electrons.}
\tablenotetext{7}{Ratio of thermal and nonthermal heating rates at 
$\Rph/5$ for cases with $\eB \le 10^{-2}$, and at the radius of
maximum $\taut$
for cases with $\eB \ge 0.1$.}
\tablenotetext{8}{Pair temperature at $\Rph/5$ for $\eB \le 10^{-2}$, and at
maximum $\taut$
for $\eB \ge 0.1$.}
\end{center}
\end{table}

\subsection{Electron/positron distribution}
\label{sec:high_energy}

Strong magnetic fields imply significant synchrotron cooling of the
high-energy $e^\pm$ pairs injected by nuclear collisions,
which can compete with Compton cooling.
Since the synchrotron photons do not create secondary pairs, 
the magnetic field has a suppressing effect on the $e^\pm$ cascade. 
The multiplicity of each subsequent pair generation in a saturated cascade is
approximately proportional to $\erad/(\erad + \eB)$,
where $\erad = \Lrad/L$ is the fraction of the jet energy carried by 
radiation. For magnetic fields approaching equipartition with radiation 
(i.e. $\eB L/L_\gamma \sim 1$), only two generations of secondary pairs 
can make a significant contribution to the total pair multiplicity. 

As a result, the total pair yield decreases from $Y \approx 0.2$ to 
about 0.005 as $\eB$ increases from 0 to 2 (Table~\ref{tbl-1}).
The pair yield is defined as $Y = {\cal M}/\gamma_0$, 
where ${\cal M}$ is the multiplicity of secondary pairs.
In the model with $\eB=2$, the pair yield is not far from the 
minimum value $Y = 1/\gamma_0$, which corresponds to ${\cal M}=1$
(no secondary pairs). The strong field completely suppresses the cascade.

\begin{figure}
\plotone{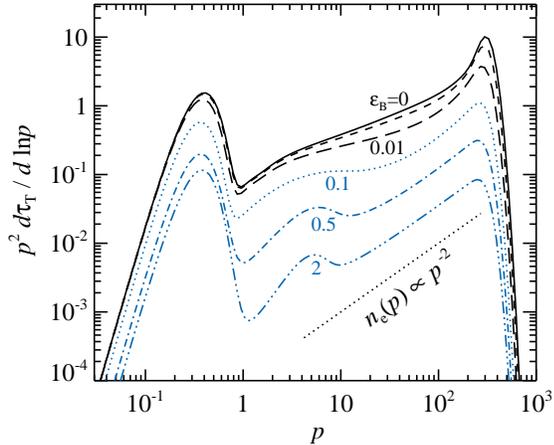}	  
\caption{
Momentum distribution of electrons and positrons at radius $r/\Rn=3$ in the 
magnetized jet. The jet parameters are the same as in 
Figures~\ref{fig:ele_distr_nonmag} and \ref{fig:GRB_nm1} except for 
magnetization. Different curves show models with different $\eB$:
$\eB = 0$ (solid), $10^{-3}$ (short-dashed), $0.01$ (long-dashed), $0.1$ 
(dotted), $0.5$ (dot-dashed) and $2$ (triple-dot-dashed).
The bump near $p\sim5$ is the result of synchrotron self-absorption,
which becomes increasingly important at high $\eB$.
The dotted line indicates the slope of a nonthermal distribution 
that would be obtained if the injected $e^\pm$ lost energy only by emitting 
synchrotron radiation, with no inverse-Compton $e^\pm$ cascade and no 
self-absorption of synchrotron radiation.
}
\label{fig:sp_mag}
\end{figure}

Figure~\ref{fig:sp_mag} shows the $e^\pm$ distribution function at 
$r=3\Rn$. The overall shape of the distribution is similar to that found
in non-magnetized jets. It has thermal and nonthermal parts. 
With increasing $\eB$, the normalization of the distribution significantly 
decreases --- the number of $e^\pm$ is reduced as the cascade is suppressed.
At $\eB\gtrsim 0.1$ the synchrotron cooling becomes dominant and 
the $e^\pm$ distribution at high energies is described by the 
power-law $\npoel(p)\propto p^{-2}$.
A new feature appears when the magnetization is high
--- the bump in the electron spectrum at $p<10$.
This bump is a result of synchrotron self-absorption that tends to 
``thermalize'' the high-energy pairs before they join the thermal 
population at low energies and thermalize via Coulomb collisions.
A similar effect is seen in the spectral simulations of 
accreting black holes in X-ray binaries and active galaxies \citep{PV09,VVP11}.

The reduction in pair production implies a smaller Thomson optical depth 
($\taut\propto Y^{1/2}$, see equation~23 in B10). As a consequence, the 
photospheric radius $\Rph$ decreases with increasing $\eB$
(Table~\ref{tbl-1}).
Another significant implication is 
the reduction in the thermal heating rate $\dot{Q}_{\rm th}$, which is 
proportional to $\taut$ (see Equation~\ref{eq:th}). Since the nonthermal 
injection rate $\dot{Q}_{\rm nth}$ remains unchanged, the ratio 
$\dot{Q}_{\rm th}/\dot{Q}_{\rm nth}$ is reduced in magnetized jets. 

\begin{figure}
\plotone{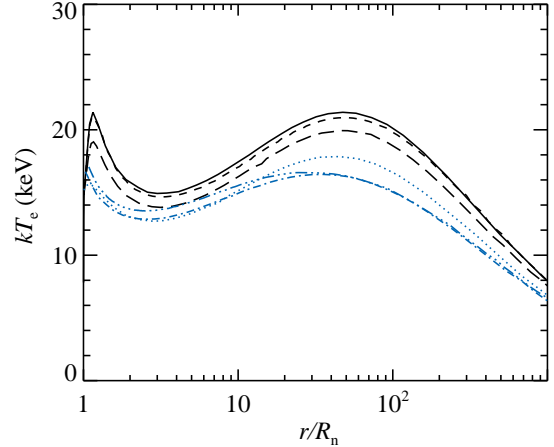}
\caption{
Temperature of the thermalized $e^\pm$ population as a function of radius, 
$\Te(r)$. The jet parameters are the same as in Figure~\ref{fig:GRB_nm1}, 
except for magnetization. Different curves show models with different $\eB$:
$\eB = 0$ (solid), $10^{-3}$ (short-dashed), $0.01$ (long-dashed), $0.1$ 
(dotted), $0.5$ (dot-dashed) and $2$ (triple-dot-dashed).
At $r/R_{\rm n}<10$ the temperature is set by the balance between Coulomb 
heating and Compton cooling; in this region $\Te(r)$ is well described by 
equations~(32) and (37) in B10. At larger radii, the contribution of 
adiabatic cooling to the thermal balance becomes non-negligible.
As a result, $\Te$ begins to slowly decrease.
}
\label{fig:ele_temp}
\end{figure}

The temperature of thermalized pairs remains remarkably stable as $\eB$ is 
increased (see Figure~\ref{fig:ele_temp}). This fact may be understood by 
noticing that
the thermal heating rate {\it per particle} does not decrease with 
increasing magnetization (the volume heating does).
Note also that the thermal pairs are unable to cool via synchrotron 
emission because of strong self-absorption. As a result their equilibrium 
temperature is weakly affected by the magnetic field.

\subsection{Radiation spectrum}

\begin{figure}
\plotone{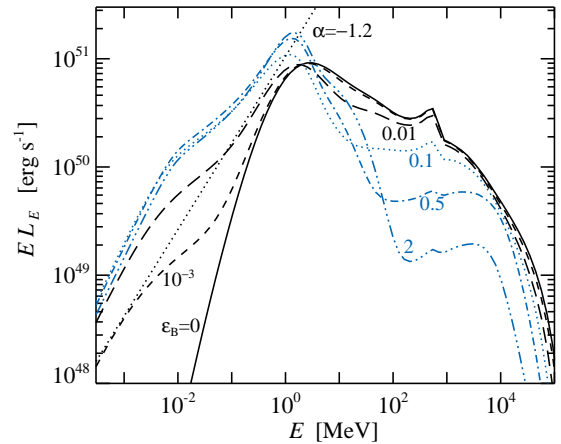}	  
\caption{
Radiation spectrum emitted by the magnetized, collisionally heated jet.
The jet parameters are the same as in Figure~\ref{fig:GRB_nm1}, except 
for magnetization.
The solid, short-dashed, long-dashed, dotted, dot-dashed and triple-dot-dashed  
curves correspond to magnetizations $\eB = 0$, $10^{-3}$, $0.01$, $0.1$, 
$0.5$ and $2$, respectively.
The straight dotted line shows a power-law spectrum with $\alpha = -1.2$.
The spectrum $L_E$ is multiplied by photon energy $E$ to make the 
differences between the models more visible in the figure.
  }
\label{fig:GRB_mag}
\end{figure}

The spectrum that should be observed from the magnetized jet is shown
in Figure~\ref{fig:GRB_mag}.
Magnetization significantly changes the radiative properties of the jet.
It suppresses the pair cascade, reduces the photospheric radius, and changes
the $e^\pm$ distribution function. It also implies a new emission component
--- synchrotron emission from the nonthermal
$e^\pm$ pairs.

Our detailed calculations confirm that the synchrotron radiation from 
strongly magnetized 
collisionally heated jets peaks at energies comparable to 1~MeV in the 
observer frame, as estimated in previous works  (\citealt{KG07}; B10).
The position of the synchrotron peak varies with magnetization as $\eB^{1/2}$.
Its amplitude increases linearly with $\eB$ for weak magnetizations 
($\eB \lesssim 0.01$) and approaches a constant value when synchrotron 
emission becomes the dominant energy loss mechanism for high-energy pairs.
The synchrotron peak remains practically buried under the 
Comptonized thermal spectrum in all models shown in Figure~\ref{fig:GRB_mag}. 
Thus, the emerging spectrum preserves the Band-type shape 
with the MeV peak. 

The presence of synchrotron radiation significantly affects the 
low-energy slope of the Band peak --- it makes the spectrum softer, i.e. 
$\alpha$ is significantly reduced compared with the non-magnetized model. 
The obtained spectral shape below 1~MeV may be described as follows.
The mixture of synchrotron and Comptonized thermal photons create 
a nearly power-law spectrum between 100 and 500~keV (with a slope 
$\alpha$ that depends on $\eB$). Below this energy range, 
the spectrum is dominated by synchrotron radiation that has a smaller,
softer slope. The resulting curvature of the spectrum may be described 
by observers as a soft excess above the power law. 
 At energies $E<10$~keV
the spectral slope changes again, as synchrotron self-absorption becomes 
important
(see Section~\ref{Sec:low}).

\subsubsection{High-energy emission}

Three mechanisms contribute to emission
at energies $E>1$~MeV:
thermal Comptonization by $e^\pm$ pairs with temperature $\sim 15$~keV
in the jet frame, nonthermal Comptonization by the $e^\pm$ cascade,
and synchrotron emission (which extends to tens of MeV when $\eB$
is large). 

Thermal Comptonization dominates at $E\gtrsim 1$~MeV, near the  
spectral peak.
The magnetized jets have approximately the same electron temperature 
$\Te$ as non-magnetized jets, and their optical depth $\taut$ is smaller.
As a result, the Kompaneets' parameter $y = 4\taut k \Te/m_{\rm e} c^2$ is 
reduced with increasing $\eB$. This leads to a steeper slope of the 
thermally Comptonized spectrum above the peak. 
Similar to the non-magnetized model, the thermally Comptonized power
law declines at $E \gtrsim 2\Gamma k\Te\sim 20$~MeV.
In models with high $\eB$, synchrotron emission
makes a comparable or even dominant contribution at these energies.
In the model with $\eB=2$, synchrotron emission contributes significantly 
to the spectrum up to 50~MeV.

The luminosity above 100~MeV is produced only by inverse Compton
scattering by the nonthermal particles. This luminosity is inevitably 
reduced with increasing $\eB$, as part of the energy of injected pairs is
lost to synchrotron emission at lower energies. The nonthermal inverse 
Compton component becomes weaker and harder with increasing $\eB$, and its 
slope approaches $\Lrad(E)\propto E^{-1/2}$. This slope is the signature of 
inverse-Compton emission with suppressed pair cascade.

Overall, the suppression of the pair cascade by synchrotron cooling destroys 
the simple power-law shape of the high-energy spectrum. Instead, a distinct 
hard component (nonthermal inverse Compton) appears above 50--100~MeV.

\subsubsection{Low-energy emission}
\label{Sec:low}

The low-energy end of the predicted spectrum is dominated by 
synchrotron emission, even when $\eB$ is small (Figure~\ref{fig:GRB_mag}).
The spectrum at energies $E<1$--10~keV is affected by 
self-absorption. It can be derived analytically as follows.

In the rest frame of the plasma, the angle-averaged synchrotron emissivity
and the absorption coefficient are given by
\citep[see e.g.][]{GS91}
\beq
  \label{magn:lowen:emiss}
  \jsyn(\nu) &  =&  \int \jsyn(\nu,p)\; \npoel(p) \; d p , \\
  \label{magn:lowen:abs}
  \kappas(\nu) & =  & -  \frac{1}{2 \me \nu^2}
  \int \jsyn(\nu,p) \: \gamma  p \: \frac{d}{dp} 
  \left[ \frac{\npoel(p)}{p^2} \right] dp.
\eeq
Here all quantities are measured in the plasma rest frame;
$\npoel(p)=n_+(p)+n_-(p)$ is the distribution function of $e^\pm$  pairs, 
and $\jsyn(\nu,p)$ is the angle-averaged synchrotron emissivity per electron. 
For analytical estimates we will use the delta-function approximation for 
the emissivity
\begin{equation}
\label{eq:syn_emis}
  4\pi\,\jsyn(\nu,p)= \frac{4}{3}\, c\, \sigmat U_{\rm B} \: 
                p^2 \, \delta\left( \nu - \gamma^2 \nuB \right),
\end{equation}
where $\nuB=eB/2\pi\me c$ is the Larmor frequency.
The synchrotron emission is produced by relativistic $e^\pm$ particles
with $\gamma\approx p$.
Then Equations~(\ref{magn:lowen:emiss})--(\ref{eq:syn_emis}) give 
\begin{equation}
\label{eq:jsyn}
  \jsyn(\nu) = \frac{\alf}{9}\,h\nuB\,p\,\npoel(p),
\end{equation}
\begin{equation}
\label{eq:ksyn}
  \frac{\jsyn(\nu)}{\kappas(\nu)} = \frac{2}{2+\sss}\,\me \nuB^2\, p^5,
\end{equation}
where $\alf=e^2/\hbar c=1/137$,
$\sss=-d\ln\npoel(p)/d\ln p$ is the local slope of the $e^\pm$ 
distribution function, and
\beq
  p\approx\gamma=\left(\frac{\nu}{\nuB}\right)^{1/2}.
\eeq

The distribution function $\npoel(p)$ in Equation~(\ref{eq:jsyn}) can be determined
by assuming a
quasi-steady flow of  $e^\pm$ particles
in the momentum space and writing
\beq
\label{eq:flow}
  \dot{p}\,\npoel(p)=\dot{n}_\pm^{\rm inj}\M(p),
\eeq
where $\dot{n}_\pm^{\rm inj}$ is the rate of particle injection at the highest 
energy $\gamma_0\approx 300$ (Equation~\ref{eq:nonth}) and $\M(p)$ is the 
multiplicity of secondary $e^\pm$ pairs created with momenta above $p$.
The synchrotron energy losses for particles emitting in the optically thin regime 
are given by
\beq
\label{eq:loss}
   \dot{\gamma}\,\me c^2\,\frac{\eB}{\erad+\eB}=\frac{4}{3}\,c\sigmat U_{\rm B} p^2.
\eeq
Using Equations~(\ref{eq:flow}) and (\ref{eq:loss}) (with $p\approx\gamma$)
the synchrotron emissivity (\ref{eq:jsyn}) becomes
\beq
\label{eq:jsyn1}
 \jsyn(\nu)=\frac{\me c^2\,\dot{n}_\pm^{\rm inj}\eB\M(\gamma)}
                 {8\pi\nuB(\erad+\eB)\gamma}.
\eeq

Let's now evaluate the range of Lorentz 
factors $\gamma>\gammas$ for particles that emit synchrotron radiation
in the optically thin regime, as a function of radius $r$.
The synchrotron photosphere can be found from the approximate condition
\beq
\label{eq:rs}
 \frac{r\kappas(\nu)}{\Gamma} = 1.
\eeq
Using Equations~(\ref{eq:ksyn}) and (\ref{eq:jsyn1}), together with the relations (\ref{eq:nonth}) and (\ref{eq:eB}), we find from Equation~(\ref{eq:rs})
\beq
  \gammas^6\sim \frac{3(2+\sss)\eB}{2^7\pi(\erad+\eB)}\,\frac{c^3}{\nuB^3}
                \,n\,\frac{\Gamma}{\Gamman}\,\taun\,\M(\gammas).
\eeq
With $[2(2+\sss)]^{1/6}\approx 2^{1/2}$, this equation gives
\beq
   \gammas \approx 
   \frac{(\me/\mprot)^{1/3}}{2^{3/4}\,\eB^{1/12}\,\Gamman^{1/3}}
     \left[\frac{\pi\M(\gammas)\,\sigman\,\Ln\Gamma^2}
                {(\erad+\eB)\sigmat L\Gamman^2}\right]^{1/6}
     \left[\frac{L r_{\rm e}}{\me c^3}\right]^{1/12}, 
\label{eq:gammas}
\eeq
where $r_{\rm e}=e^2/\me c^2\approx 2.82\times 10^{-13}$~cm is the classical 
electron radius. This gives $\gammas\approx 10$ for the parameters adopted in our models.
Note that $\gammas$ does not depend on $r$ and weakly depends on the parameters of the jet.

Synchrotron radiation at a given radius $r$ is self-absorbed at energies 
$E<\Es$ where
\beq
  \label{eq:Es}
  \Es(r)\sim \Gamma\gammas^2\,h\nuB\sim \frac{\hbar e}{\me c}\,
         \frac{\gammas^2}{r}\,\left(\frac{2\eB L}{c}\right)^{1/2}.
\eeq
The entire heating region $r>\Rn$ is transparent to synchrotron absorption 
for photons above $\Es(\Rn)$. For the jet models calculated in this
paper $\Es(\Rn)\sim 1$--$10$~keV.
Observed radiation at lower energies $E$ comes mainly from the corresponding 
synchrotron photosphere $\rphs(E)$
that can be found by expressing $r$ from Equation~(\ref{eq:Es}).

Neglecting the self-absorbed radiation from $r<\rphs(E)$, we estimate the 
observed spectral luminosity $\Ls(E)$ by integrating the synchrotron 
emissivity over the optically thin region $r>\rphs(E)$,
\begin{equation} \label{magn:lowen:eps}
  \Ls(E) \approx \int_{\rphs(E)}^{\infty} \,\frac{4\pi\jsyn(\nu)}{h}\,
                       4\pi r^2\, dr.
\end{equation}
Here we approximated $h\nu\approx E/\Gamma$ and used the fact that the 
angle-integrated emissivity $4\pi\jsyn$ is the same in the static and 
comoving frames within a factor $C\approx 1$.\footnote{
      This factor is given by 
      \bez 
        C = \frac{1}{2}  \int_{-1}^1  
           \left(\frac{\cD}{\Gamma}\right)^{\frac{\delta-1}{2}} d\mu 
        \approx \frac{1}{2}  \int_{-1}^1  (1+\mu)^{\frac{\delta-1}{2}} d\mu 
         = \frac{2^{\frac{\delta+1}{2}} }{\delta+1}\approx 1. 
      \eez } 
Using the slope of $e^\pm$ distribution $\sss\approx 2$, one can show that 
$\jsyn(\nu)r^2$ scales with radius as $r^{-(1+\sss)/2}=r^{-3/2}$ and 
its integral peaks near $\rphs(E)$.
Substituting $\jsyn(\nu)$ from Equation~(\ref{eq:jsyn1}),
using the relations $\gamma^2 \sim E/\Gamma h\nuB$, $\Gamma^2 B^2 r^2=2\eB L/c$ and Equation~(\ref{eq:nonth}),
we obtain
\beq
\label{Ls}
 \frac{\Ls(E)}{L} \approx \frac{\alf}{\me c^2}\,\Gamma^{-1}\,\gammas^5\,\eB, 
                              \qquad E<\Es(\Rn),
\eeq 
where $\gammas$ is given by Equation~(\ref{eq:gammas}).
The emission at any photon energy $E<\Es(\Rn)$ peaks near $\rphs(E)$ 
and is produced by $e^\pm$ particles with the same $\gammas\sim 10$.
Note that $\Ls(E)=const$
(flat spectrum), which corresponds to photon index $\alpha=-1$. 
This fact is a consequence of $B\propto r^{-1}$ and $\npoel(p)\propto r^{-2}$.
For a similar reason, a flat spectrum was derived for opaque radio 
jets in AGN \citep{BK79}. 
Equation~(\ref{Ls}) is in approximate agreement with our numerical results 
(Figure~\ref{fig:GRB_mag}).

\subsection{Radiative efficiency}

\begin{figure}
\plotone{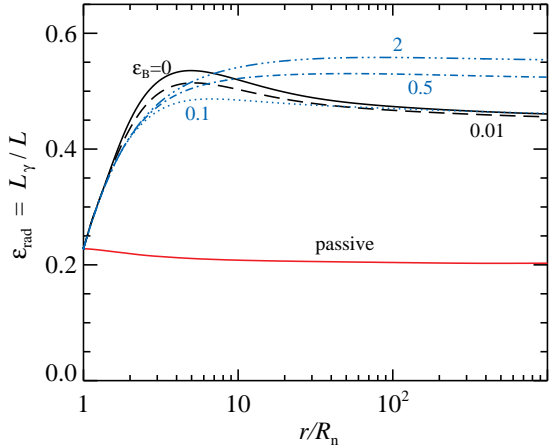}	 
\caption{Fraction of the flow energy carried by radiation
(radiative efficiency)
as a function of distance from the central source, for different magnetizations: $\eB = 0$ 
(solid line), $0.01$ (long-dashed), $0.1$ (dotted), 
$0.5$ (dot-dashed) and $2$ (triple-dot-dashed). For comparison, the radiative 
efficiency for a passively cooling flow is also plotted (lower solid line). 
}
\label{fig:GRB_eff}
\end{figure}

The radiative efficiency of the jet $\erad$ is defined as the ratio of the 
photon luminosity $L_\gamma$ to the kinetic luminosity of the plasma 
outflow $L$. Note that $L(r)\approx const$ for the matter-dominated jet 
and $L_\gamma$ evolves with radius. This evolution is shown in 
Figure~\ref{fig:GRB_eff}. The final efficiency is given by the asymptotic 
value of $\erad=L_\gamma/L$ at large radii.
 
The photon luminosity prior to the onset of dissipation ($r=R_{\rm n}$) is 
found from the passively cooling jet model.
One can see from Figure~\ref{fig:GRB_eff} that $L_\gamma$
is greatly increased by the intense collisional heating at 
$r \gtrsim R_{\rm n}$. The net energy given to radiation by collisional 
heating outside a given radius is obtained by integrating 
Equations~(\ref{eq:nonth}) and (\ref{eq:th}) over 
volume; the result is proportional to $r^{-1}$.  
Thus, heating peaks near $R_{\rm n}$   and
continues with a smaller rate at larger radii.
On the other hand, $L_\gamma$ is reduced by 
adiabatic cooling, in particular in the opaque zone $r<\Rph$. 
The competition between collisional heating and adiabatic cooling shapes
$L_\gamma(r)$. In weakly magnetized jets, the photospheric radius is large, 
$\Rph\sim 20R_{\rm n}$, and adiabatic cooling is more efficient. 
It begins to win over collisional heating at $r\sim 5R_{\rm n}$ and 
somewhat reduces $L_\gamma$. In strongly magnetized jets, the photospheric 
radius $\Rph$ is smaller because the magnetic field suppresses the 
production of $e^\pm$ pairs (Section~\ref{sec:high_energy}). In this case, 
adiabatic cooling is less efficient.

In all cases shown in Figure \ref{fig:GRB_eff} the final $\erad$ is close 
to 50\%. We conclude that the radiative efficiency of collisionally heated 
jets remains high for the entire range of $\eB$ considered in this paper.


\section{Discussion}		\label{GRB:conclusions}

Only two radiative processes are in principle capable of producing
the GRB spectrum with an extended high-energy tail:
synchrotron emission and Comptonization. 
These are two basic modes of energy transfer from a heated plasma to 
radiation, and their theory was developed long ago. In particular, 
the formation of photospheric spectra through Comptonization in X-ray 
sources was investigated in the 1970s, with various applications from 
accretion disks to the expanding Universe \citep[see e.g.][]{PSS83}.
Comptonization in GRBs has been discussed for about two decades (see 
Section~\ref{sec:comp_spec} below).
Besides the fact that subphotospheric heating creates a nonthermal-looking 
radiation spectrum, one would like to know what the heating process is.
The answer to
this question can help disentangle GRB physics --- the nature 
and composition of the jet and the central 
engine.\footnote{The situation may be compared with X-ray binaries, where 
  Comptonization is thought to occur in a corona of the accretion disk or the inner hot flow, and 
  the main puzzle is how the plasma is heated.}
Progress in this direction was hampered for many years by the complexity 
of collisionless processes that were usually invoked in GRB production.

GRB spectra can be shaped by three possible heating mechanisms:
(1) collisionless shocks, (2) magnetic dissipation, and 
(3) collisional dissipation. 
Least understood is magnetic dissipation, and it is usually modeled by 
introducing phenomenological parameters. A similar phenomenological approach 
was used for internal shocks in the jet until recent numerical simulations 
began to provide insights into shock physics. 
Collisional dissipation is the most straightforward mechanism of these
three. It can be calculated exactly, from first principles. B10 recently 
showed that collisional heating possesses the key features of a successful 
GRB mechanism: efficient electron heating, high radiative efficiency and, 
most importantly, it generates the observed Band-type spectrum.

In this paper, we   calculated the emission from collisionally heated 
jets by solving the time-dependent, coupled kinetic 
equations for the particle and photon distributions inside the jet.
The advantage of our numerical code is the accurate modeling of all 
relevant kinetic and radiative processes including Coulomb collisions 
and synchrotron self-absorption. It allowed us to systematically explore
the effects of jet magnetization on the emerging spectrum. 
We calculated the emerging radiation using the equation of radiative 
transfer (B11) instead of the Monte-Carlo method used by B10. The only 
disadvantage of our code is that it uses angle-averaged 
coefficients of emission and absorption in the transfer equation. 
Comparison with exact transfer models of B10 and B11 shows that this
approximation is reasonable, in particular at photon energies below GeV.
Note also that this is the first implementation of radiative transfer 
calculations in a kinetic GRB code. The previously developed kinetic 
codes assumed isotropy of radiation in the rest-frame of the plasma 
(\citealt{PW05}; VP09), which is invalid in the main region of interest $\taut<10$.

 \subsection{Spectra from non-magnetized and magnetized jets}

First, we calculated the emission from the fiducial baryonic jet model of 
B10 with zero magnetic field ($\eB=0$). Our results agree with B10: 
the spectrum peaks at 1~MeV, has the low-energy photon index 
$\alpha\approx 0.4$ and the high-energy photon index $\beta\approx -2.5$.
We conclude that the Band-type spectrum is a robust prediction of the 
collisional-heating model. 
Similar to B10, we stress the {\it ab initio} character of our calculations: 
the dissipation process and generated radiation are derived from first 
principles, without introducing any phenomenological parameters for the 
heating mechanism.

The collisional heating peaks below the photosphere, where the 
$\gamma$-$\gamma$ opacity is large for multi-GeV photons emitted
in the $e^\pm$ cascade.
The nuclear collisions also occur at much larger 
radii (although with a smaller rate) where the $\gamma$-$\gamma$ optical 
depth is reduced and the high-energy photons can escape. As a result, 
instead of a cutoff at a few GeV, the predicted spectrum
exhibits a moderate steepening at $\sim 5$~GeV 
and extends up to $\sim 100$~GeV (Figure~\ref{fig:GRB_nm1}). 
A similar behavior is 
seen in the Monte-Carlo results of B10, with somewhat smaller 
normalization of the multi-GeV emission. This difference must be 
caused by the angle-averaged approximation for $j_\nu$ and $\kappa_\nu$
that was adopted in our calculations of radiative transfer. As discussed
above, the accuracy of this approximation is reduced at photon energies 
$E>1$~GeV.

Then we calculated the emission from magnetized jets.
We considered the range of magnetization parameters 
$0<\eB<2$,   so the magnetic energy was at most comparable to the energy of 
the baryonic component.
We did not consider magnetically dominated jets with $\eB\gg 1$
for two reasons. First, the strong magnetic field can change the jet 
dynamics. Then the coasting approximation (which is reasonable for a 
matter-dominated jet) would be invalid. Second, the strongly 
magnetized case is computationally more difficult. The generated 
spectra in this regime are left for a future study. 
 
Our calculations show that magnetization has a small effect on the 
position of the spectral peak $\Epeak$ --- it is only slightly shifted to 
lower energies. The spectrum still has the Band-type shape around the peak, 
with $\alpha$ and $\beta$ depending on $\eB$ (Figure~\ref{fig:GRB_mag} and 
Table~\ref{tbl-1}). Magnetization significantly affects the shape of the 
spectrum at high ($E\gg\Epeak$) and low ($E\ll\Epeak$) energies.

The high-energy emission is reduced with increasing $\eB$ 
(Figure~\ref{fig:GRB_mag}), because the nonthermal $e^\pm$ generated by 
nuclear collisions emit less via inverse Compton scattering and
more at low energies via the synchrotron mechanism.
This change is significant if $\eB\gg 10^{-2}$.
Synchrotron cooling of the injected high-energy $e^\pm$ 
has a throttling effect on the pair cascades. 
As a result, magnetized jets have
smaller densities of $e^\pm$ pairs
and smaller photospheric radii $\Rph$.

The low-energy emission is increased with increasing $\eB$.
An important effect of magnetization is the softening of the spectral 
slope below $\Epeak$.
This is the result of synchrotron emission from nonthermal pairs. 
At energies below $\sim 30$~keV a clear soft excess 
is predicted, even for small $\eB\sim 10^{-3}$. 
Our model also predicts that the spectrum below a few keV should have  
the photon index $-1$. This spectrum extends down to the optical band. 

Note that  photon spectra in this paper are not corrected for the cosmological redshift $z$.
The actually observed energy of all photons is smaller by the factor of $(1+z)^{-1}$. 
In particular, the soft excess is redshifted to $\sim 10$~keV for a typical $z$ of GRBs.

\subsection{Comparison with other models of subphotopsheric Comptonization}
\label{sec:comp_spec}

\citet{Thompson94} proposed that GRB photons are Comptonized in the 
subphotospheric region by turbulent bulk motions of the plasma in the jet.
His model assumed that the jet is dominated by magnetic field
and pictured Alfv\'enic turbulence generated by some instabilities 
(e.g. by reconnection). Building a detailed theory of turbulence from 
first principles would be a formidable task, and \citet{Thompson94} made 
estimates assuming that the plasma motions are limited by radiation drag. 
Then the bulk Comptonization occurs in the unsaturated regime and controlled 
by the efficiency of turbulence generation against the drag.
The predicted GRB spectrum is sensitive to this unknown efficiency,
which may vary with radius.
The mechanism is very different from collisional heating. Direct comparison
of the model predictions is difficult, because \citet{Thompson94} used the
theory of Comptonization in static sources, which
is not valid for GRBs (see discussion in Section~\ref{sec:method}).
Like in the expanding Universe, the number of scatterings in a GRB jet scales as $\sim\taut$
(not as $\taut^2$) and is controlled by the decrease in density,
not the escape of photons from the plasma.
The expected spectrum can only be found by
solving radiative transfer in the expanding jet, and we emphasize the 
importance of accurate transfer calculations for GRB models.\footnote{
     When heating occurs in a static source, the emerging 
     spectrum can have two peaks, resembling the Comptonized spectra of 
     accretion-disk corona. In contrast, the spectrum emitted by the GRB jet 
     has one peak.}

\citet{MR00b} and \citet{ReesMeszaros05} discussed subphotospheric 
Comptonization in general terms and highlighted its importance for GRBs.
They argued that Comptonization helps explain observations, although
they did not consider any concrete model.
More detailed calculations were performed by 
\citet{Peer06}. They still did not focus on any concrete physical 
scenario and experimented with various phenomenological models, using
free parameters to describe the heating process:
the amount of dissipated energy and its fraction that is given to electrons, 
the shape of the electron distribution, and the optical depth at which the 
dissipation occurs. Various combinations of these parameters led to various 
spectra. In contrast, we investigated the concrete physical model of collisional 
dissipation, which predicts the thermal and nonthermal heating rates, 
both scaling as $r^{-1}$.
It is therefore hard to compare their results with ours. 
We can only compare the technical tools. \citet{Peer06} used the 
kinetic code developed by \citet{PW05}. In many respects, their
code is similar to ours.
Their code assumes, however,
that radiation is isotropic in the comoving frame of the jet. We gave up this 
approximation because 
it is violated even below the photosphere
(see Section~\ref{sec:method}) and 
instead solved the radiative transfer equation. Note also that 
\citet{Peer06} argued that GRB jets never become dominated by $e^\pm$ 
pairs. In contrast, we find that jets with magnetization $\eB\ll 1$ 
have a large pair loading factor $\npoel/n\sim 20-40$, which changes the
photospheric radius $\Rph$
by the factor of 20--40. 

\citet{Giannios06} investigated Comptonization in jets that carry 
alternating magnetic fields and are heated by magnetic reconnection
\citep{DS02}. This model has two phenomenological parameters
--- the reconnection rate and the fraction of dissipated energy that is 
deposited into electrons. The effect of magnetic dissipation on the electron 
distribution is unknown. \citet{Giannios06} assumed a thermal electron 
distribution and studied the radiation produced by this model.
He used a Monte-Carlo code and an iterative technique to simulate the
radiative transfer in the jet.\footnote{The numerical
     methods used by \citet{Giannios06} and B10 are similar, except that 
     the iterations of 
     the electron temperature in \citet{Giannios06} are simplified: the 
     profile $\Te(r)$ is searched in the power-law form, with two parameters 
     to iterate --- the normalization and the slope of the power law.}
We note that 
the jet model in \citet{Giannios06} is significantly different from B10 
and our work. In particular, the jet continues to accelerate as
$\Gamma\propto r^{1/3}$ in the heated subphotospheric region while we
focus on matter-dominated jets with $\Gamma\approx const$.
The electron temperature scales with radius approximately as 
$\Te\propto r^{5/3}$ while in our model we find that $\Te$ varies only by a factor of $\sim 2$ for three decades
in radius (Figure~\ref{fig:ele_temp}). Note also that \citet{Giannios06} assumed pure thermal heating, leading to 
negligible pair creation. 
In contrast, our work and B10 predict a nonthermal $e^\pm$ cascade, 
which has significant
effects on the observed spectrum at high energies 
(above 10~MeV) and at low energies where synchrotron emission from the 
cascade becomes dominant.

\citet{Lazzati10} considered a simplified model for Comptonization 
following an impulsive heating event below the photosphere. In their model, 
all electrons in a slab of optical depth $\taut\sim 2$ are suddenly given a Lorentz factor 
$\gamma$ and then cooled by blackbody radiation.
This setup can produce a variety of photon spectra depending on 
photon-to-electron ratio $n_\gamma/n_e$, $\taut$, and $\gamma$. 
It has, however, a problem. The sudden heating of the slab (faster than 
Compton cooling) can hardly describe any physical dissipation mechanism, 
as different parts of the slab are causally disconnected. The causal contact 
(light-crossing) time for a region with $\taut\sim 2$ is orders of magnitude 
longer than Compton cooling time.

\subsection{Comparison with observed GRB spectra}

The main part of observed GRB radiation is emitted with a Band-type 
spectrum --- a smoothly broken power law that peaks near $0.3$~MeV. 
The spectral slopes below and above the peak ($\alpha$ and $\beta$) vary \citep{Preece00}.
The most frequently measured low-energy slope $\alpha$ is close to $-1$,
but in some cases it reaches $1$ (Rayleigh-Jeans slope of the Planck
spectrum). The most frequently measured high-energy slope $\beta$ is near 
$-2.5$; it also significantly varies from burst to burst, roughly between 
$-2$ and $-3$.

The spectra predicted by our model appear to agree with observations.
The predicted spectral peak is close to MeV/$(1+z)$.
The average observed values of $\alpha\sim -1$ and $\beta\sim -2.5$ are 
consistent with the collisionally heated jet with magnetization 
$\eB\sim 10^{-3}-10^{-2}$. A lower magnetization $\eB<10^{-3}$ 
leads to harder $\alpha$ up to 0.4. Larger slopes are not expected
from the matter-dominated jets. The slope $\alpha=1$ is predicted only in the 
radiation-dominated regime -- in this case the jet emits a Planckian spectrum 
(B11).

It should be noted that our calculations 
give practically {\it instantaneous} observed spectra. The main part of the 
prompt emission (photon energies from soft X-rays to a few GeV) is dominated 
by the photosphere in our model. The emission is produced by a sequence of 
short-lived (in observer time) independent emitters passing through $\Rph$ 
--- the sequence of shells of thickness $\delta r\lesssim \Rph/\Gp^2$, 
which corresponds to duration $\delta t_{\rm obs}\sim 10^{-4}$~s for 
typical parameters of GRB jets. The spectra obtained by modern detectors 
do not provide this high temporal resolution, because of insufficient 
photon statistics. The observed spectra must be a mixture of emissions from 
different shells, possibly with significant variations in $L$ and $\Epeak$. 
The superposition of different instantaneous spectra generally tends 
to reduce the observed slope $\alpha$
and can give $\alpha\sim -1$ even for non-magnetized jets 
(R.~Mochkovitch, in preparation).

An interesting feature of the magnetized model of collisionally heated
jet is the soft excess that appears at energies $\sim 30(1+z)^{-1}$~keV.
It may provide an explanation for the X-ray excesses that have been 
observed in several bursts \citep{Preece96}, including the recent example 
of GRB~090902B \citep{Abdo_090902B}.

Prompt optical emission is expected from collisionally heated jets,
with a slope $\Lrad(E)=const$ (photon index $-1$). 
This emission peaks at the self-absorption radius $\rphs(E)\gg \Rph$ and 
should be observed with a time lag $\sim \rphs/c\Gamma^2\sim 1$~s with 
respect to the MeV radiation that is released at $r\sim\Rph$.
The prompt optical emission has been detected in some GRBs, but the 
data are sparse, as prompt optical observations are difficult. 
In some cases (e.g. GRB~080319B) the optical luminosity appears to 
exceed the predictions of our model, suggesting an additional 
source of emission (e.g. shocks or neutron decay, see B10).
  
The high-energy index $\beta$ predicted by the collisional model
can easily be steeper than $-2.5$, because of a large $\eB$
(Figure~\ref{fig:GRB_mag}) or less efficient heating (B10). 
The model can also accommodate very hard slopes $\beta>-2$ that are observed 
in some GRBs (see e.g. Figure~7 in B10).

Our model predicts significant multi-GeV emission, especially when 
$\eB$ is small. It implies that the prompt GRB emission can
contribute to the radiation observed by LAT instrument of the 
{\it Fermi} telescope. The prompt high-energy photons should mix together 
with radiation from the other, long-lived, high-energy source that was 
identified by LAT in a fraction of GRBs and whose origin is debated.

Overall, our results support the view that the main peak of GRB spectra 
is dominated by photospheric radiation of the ultra-relativistic jet from 
the central engine. Note that photospheric emission is often erroneously 
pictured as a Planckian component in GRB spectra. As shown in B11,
the photospheric spectrum has a Planckian shape only if the jet energy is 
strongly dominated by radiation at $\Rph$. This radiation-dominated regime 
requires an extremely high Lorentz factor, which corresponds to a very 
baryon-poor jet. 
Baryon-dominated jets cannot produce Planckian spectra, even if the jet
experiences no dissipation and
only emits adiabatically cooled thermal photons from the photosphere
(Figure~1 in B10). In reality, the jet should experience strong collisional dissipation,
which has a huge impact on its
photospheric radiation and makes it even more different from Planckian. 
The only feature shared with Planckian radiation is the single spectral peak 
at $\Epeak\sim 1$~MeV. The photospheric spectrum of the collisionally 
heated jet is much broader than Planckian and fits the observed GRBs.


\begin{acknowledgements}
This work was  supported by the Wihuri foundation and ERC Advanced Research 
Grant 227634 (IV), NSF grant AST-1008334 and NASA grant NNX10AO58G (AMB), 
and the Academy of Finland grant 127512 (JP). 
\end{acknowledgements}


\appendix

\section{Radiative transfer}
\label{append:RTE}
 
Equation~(\ref{method:ph_eq}) is valid for relativistic outflows in the 
limit $\Gamma\rightarrow\infty$. For finite $\Gamma\gg 1$, the equation 
accurately describes the transfer in a steady jet. Its accuracy is reduced
if the jet is variable on scales $\delta r\lesssim r/\Gamma^2$, as 
discussed in B11. Here we give a more formal discussion of transfer in 
variable jets and the applicability of the steady 
equation~(\ref{method:ph_eq}).
The general equation of radiative transfer in a spherically symmetric 
outflow with any Lorentz factor $\Gamma(t,r)$
can be written as \citep{Mihalas80}
\begin{align}
  \widehat{\cal L} I_{\nu} = 
   \left[ \frac{\Gamma\Beta}{r} (1-\mu^2) + \frac{\mu}{\Beta\Gamma} \; 
         \widehat{\cal L} \Gamma \right] \; 
  \nu^3 \frac{\partial}{\partial\ln{\nu}} \left( \frac{I_{\nu}}{\nu^3} \right)
- (1-\mu^2) \left[ \frac{\Gamma}{r} (1+\Beta\mu) - \frac{1}{\Beta\Gamma} \; 
   \widehat{\cal L} \Gamma \right] \; \frac{\partial I_{\nu}}{\partial\mu}
+ j_{\nu} - \kappa_{\nu} \; I_{\nu},			\label{app:RTE}
\end{align}
where $\widehat{\cal L}$ is the differential operator
\begin{align}
  \widehat{\cal L} = \Gamma(1+\Beta\mu) \, \frac{\partial}{c\partial t}
  + \Gamma(\mu + \Beta) \, \frac{\partial}{\partial r}.	     \label{app:L}
\end{align}
  Here $\Beta=(1-\Gamma^{-2})^{1/2}$ is the flow
velocity in units of $c$, $t$ and $r$ are time and radial 
coordinate measured in the static frame, whereas the cosine $\mu$ of the 
photon angle relative to the radial direction, the specific intensity 
$I_{\nu}$, the emission and absorption coefficients $j_{\nu}$ and 
$\kappa_{\nu}$ are all measured in the local comoving frame.

Let  us define a coordinate $s$ by the following implicit relation
\begin{align}
   s = c t - \frac{r}{\Beta(s)}.	\label{app:transf}
\end{align}
For a matter-dominated (coasting) flow, $\Gamma(s)\approx const$ and
$s$ can serve as a Lagrangian coordinate that labels different shells 
of the flow, as $s=const$ along the worldlines of the shells. Changing 
variables $r,t \rightarrow r,s$, we rewrite   the operator (\ref{app:L}) as
\begin{align}
  \widehat{\cal L} = \Gamma(\mu + \Beta) \; 
    \frac{\partial}{\partial r} - \frac{\mu}{\Beta\Gamma} 
  \left[ 1 - \frac{r}{\Beta^{\,3}\Gamma^2}\frac{d\ln{\Gamma}}{ds} \right]^{-1} 
    \; \frac{\partial}{\partial s}.                  \label{app:Lrs}
\end{align}
Here the spatial derivative $\partial/\partial r$ is taken along the worldline
(i.e. at $s=const$). Note that the flow worldlines do not cross (i.e. 
there are no caustics) as long as the bracket in Equation (\ref{app:Lrs}) is 
non-zero.
 
If the flow varies on scales $\delta r\sim r/\Gamma^2$, the two parts
of the operator~(\ref{app:Lrs}) are comparable. The problem simplifies 
(becomes formally equivalent to the steady problem) if the flow varies 
on scales much larger than $r/\Gamma^2$. Then the operator 
$\widehat{\cal L}$ is dominated by the first term that is
proportional to $\partial/\partial r$. The left-hand side of 
Equation~(\ref{app:RTE}) simplifies to $\widehat{\cal L} I_{\nu} = 
\Gamma(\mu +\Beta)\;\partial I_{\nu}/\partial r$. Similarly, the 
right-hand side of Equation~(\ref{app:RTE}) simplifies and
the transfer equation becomes,
\begin{align}
\frac{1}{r^2}\frac{\partial}{\partial \ln{r}} \left[
   (\mu+\Beta) \; r^2 I_{\nu} \right] = \frac{r}{\Gamma}  
       \left( j_{\nu} - \kappa_{\nu} \; I_{\nu} \right)
 + \Beta (1-\mu^2)  \, \frac{\partial I_{\nu}}{\partial\ln{\nu}}
 -\frac{\partial}{\partial\mu} \left[ (1-\mu^2)(1+\Beta\mu)\; I_{\nu} \right],
\label{app:RTE2}
\end{align}
which becomes Equation~(\ref{method:ph_eq}) for outflows with $\Beta\approx 1$.
Formally, this is achieved 
simply by replacing $\partial/\partial r|_t$ by the derivative along the 
flow worldline $\partial/\partial r|_s$. Equation~(\ref{app:RTE2}) can 
also be written using the flow proper time as the independent variable, 
$t=r/c\Beta\Gamma$.


\end{document}